\documentclass[12pt]{article}

\usepackage{amssymb}
\usepackage{amsmath}

\usepackage{epsfig}

\begin{document}%



\title{\bf Can universality of the QCD evolution be checked in W boson
decays into hadrons?}


\author{A.V. Kisselev\thanks{E-mail: alexandre.kisselev@ihep.ru} \
and V.A. Petrov\thanks{E-mail: vladimir.petrov@ihep.ru} \\
\small Institute for High Energy Physics, 142281 Protvino, Russia}

\date{}

\maketitle

\thispagestyle{empty}

\bigskip

\begin{abstract}
Hadron multiplicity from $W$ boson is calculated in pQCD. The
agreement of our theoretical predictions with the LEP data says in
favor of universality of the QCD evolution in hard processes.
\end{abstract}



\section{Introduction}

Experiments at LEP and SLAC colliders have shown that the multiple
production of hadrons in $e^+e^-$ annihilation depend on the mass
of the primary (anti)quarks which launch the process of the QCD
evolution. Let us consider a heavy quark induced event,
\begin{equation}\label{e-e+_QQ}
e^+e^- \rightarrow Q \,\bar{Q} \rightarrow X \;,
\end{equation}
where $Q$ means a heavy ($c$ or $b$) quark, and an $e^+e^-$ event
induced by the pair of light quarks:
\begin{equation}\label{e-e+_ll} e^+e^- \rightarrow
l\,\bar{l} \rightarrow X \;.
\end{equation}
Here and in what follows $l$ denotes $u,\ d$ or $s$-quarks which
are assumed to be massless. In both cases, $X$ means a system of
final hadrons.

Let $N_{QQ}(W^2,m_Q^2)$ and $N_{ll}(W^2)$ be the average
multiplicities of charged hadrons in $e^+e^-$ events with heavy
\eqref{e-e+_QQ} and light primary quarks \eqref{e-e+_ll},
respectively. $W$ is the invariant energy of colliding leptons,
$m_Q$ is the mass of the (anti)quark $Q$. It appeared that
differences between the light and heavy quark induced
multiplicities,
\begin{equation}\label{delta_Ol_def}
\delta_{Ql} = N_{QQ} - N_{ll} \;,
\end{equation}
become independent of the collision energy $W$, but depend only on
the heavy quark mass $m_Q$. QCD calculations describe the
phenomenon quite well~\cite{Petrov:95}-\cite{Kisselev:08} (see
also \cite{Dokshitzer:06}).

The QCD calculations of the hadron multiplicities in $e^+e^-$
events, in particular, hadron multiplicity from $Z$ boson, are in
a good agreement with the data. Their energy dependence is defined
by the QCD evolution of the parton showers. The aim of the present
paper is to calculate the hadron multiplicity from the $W$ boson
in pQCD, and thus to check once more the universality of the QCD
evolution in hard processes.



\section{Hadron multiplicity from the $\mathbf{W}$ boson}

From now on we will consider the multiple hadron production in
$e^+e^-$ annihilation mediated by production of a pair of $W$
bosons. We will refer to $e^+e^-$ event as the heavy quark event
if one of the $W$ bosons produces a lepton pair, while the other
decays into hadrons via charm production, for instance:
\begin{equation}\label{WW_Ql}
e^+e^- \rightarrow W^+ W^-, \qquad W^+ \rightarrow c \, \bar{s}
\rightarrow X, \quad W^- \rightarrow \mu^-\bar{ \nu}_{\mu}, \;.
\end{equation}
In the light quark event final hadrons are fragments of the light
quark-antiquark pair produced by one of the $W$ bosons:
\begin{equation}\label{WW_ll} e^+e^-  \rightarrow W^+
W^-, \qquad W^+ \rightarrow l\,\bar{l'} \rightarrow X, \quad W^-
\rightarrow W^- \rightarrow \mu^- \bar{\nu}_{\mu}, \;.
\end{equation}

Let us define the average multiplicities of charged hadrons in the
above mentioned processes as  $N_{Ql}(W^2,m^2)$ and $N_{l}(W^2)$,
where $m$ is the mass of the heavy (charm) quark.%
\footnote{For generality, we will often use the notation $Q$ in
our formulae, having in mind that $Q$ means charm quark in the
most of the cases ($Q=c$).}
Our main goal is to establish a relation between these two
multiplicities. Namely, we will calculate the difference
\begin{equation}\label{Delta_Ol_def}
\Delta_{Ql} = N_{Ql} - N_{l} \;,
\end{equation}
analogous to quantity \eqref{delta_Ol_def}. Let us note that the
multiplicity $N_{l}$ is equal to the multiplicity in the $e^+e^-$
event taken at the energy $W=m_W$, i.e. $N_{l}$ = $N_{ll} (W^2 =
m_W^2)$, where $m_W$ is the $W$ boson mass.

\begin{figure}[htb]
\begin{center}
\epsfysize=4cm \epsffile{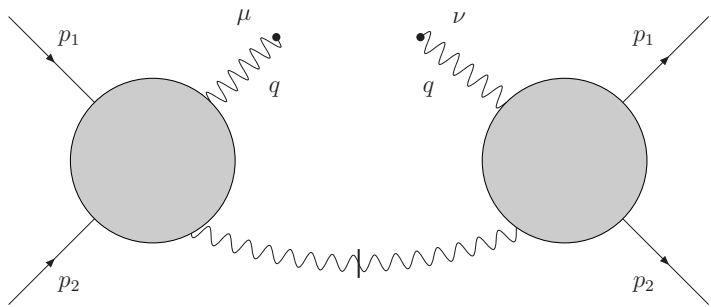}
\end{center}
\caption{The diagram describing the process $e^+(p_1) + e^-(p_2)
\rightarrow W^{\pm}(q) + W^{\mp}(p_1 + p_2 - q)$, where $\mu$,
$\nu$ are Lorentz indices of the $W$ boson (wavy lines). The cut
line of the $W$ boson means that it is on-shell particle.}
\label{fig:W_emission}
\end{figure}

\begin{figure}[htb]
\begin{center}
\epsfysize=4cm \epsffile{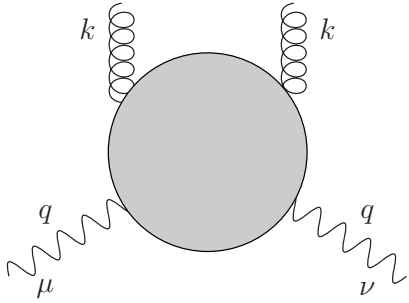}
\end{center}
\caption{The inclusive distribution of the ``massive'' gluon jet
with the 4-momentum $k$ (spiral line). The wavy line is the $W$
boson, whose 4-momentum is $q$, and Lorentz indices are $\mu$,
$\nu$.}
\label{fig:gluon_W}
\end{figure}

Hadron multiplicity in a heavy quark event is represented by the
following equation:
\begin{equation}\label{mult_Ql}
N_{Ql}(m_W^2,m^2) = (n_Q +  n_l) + \tilde{N}_{Ql}(m_W^2,m^2) \;,
\end{equation}
with
\begin{align}\label{mult_0}
\tilde{N}_{Ql}(m_W^2,m^2) &= \frac{1}{N_0} \, \int \!
\frac{d^4q}{(2\pi)^2} \, \delta (q^2 - m_W^2) \, \delta ((p_1 +
p_2 - q)^2 - m_W^2)
\nonumber \\
&\times \Pi_{\mu \nu}(p_1, p_2, q) \! \int \! \frac{d^4k}{(2\pi)^4
\, k^2} \, \Phi^{\mu \nu}(q, k, m) \, n_g(k^2) \;,
\end{align}
where we have assumed that both bosons are on-shell.%
\footnote{An account of the $W$ boson distribution in its
invariant mass $q^2$ will be discussed below (see
Eqs.~\eqref{W_width}-\eqref{n_W_corr_num}).}
Two first terms in the r.h.s. of Eq.~\eqref{mult_Ql}, are the
multiplicities from the leading (anti)quark $Q$ and $l$. They are
known from the data. The tensor $\Pi_{\mu \nu}(p_1, p_2, q)$
describes the process $e^+(p_1) \ + \ e^-(p_2) \rightarrow
W^{\pm}(q) \ + \ W^{\mp}(p_1 + p_2 - q)$, where $\mu$, $\nu$ are
the Lorentz indices of the W boson with the 4-momentum $q$ (see
Fig.~\ref{fig:W_emission}). Two blobs in Fig.~\ref{fig:W_emission}
are tree diagrams with the $\gamma/Z^0$ exchange in the
$s$-channel, and electronic neutrino exchange in the $t$-channel.
The tensor $\Phi^{\mu \nu}(q, k, m)$ describes the emission of the
gluon jet with the 4-momentum $k$ produced by this $W$ boson (see
Fig.~\ref{fig:gluon_W}).
The quantity $n_g(k^2)$ is the average multiplicity of the hadrons
in the gluon jet with the invariant mass
$k^2$~\cite{Petrov:95,Kisselev:07}. $N_0$ is the normalization
factor.

It is convenient to use Lorentz gauge in which the tensor part of
the W boson propagator has the form:
\begin{equation}\label{propagator}
d_{\mu \nu}(q) =  - g_{\mu \nu} + \frac{q_{\mu} q_{\nu}}{q^2} \;,
\end{equation}
then
\begin{equation}\label{partonometer_trans}
q^{\mu} \Pi_{\mu \nu}(p_1, p_2, q) = q^{\nu}\Pi_{\mu \nu}(p_1,
p_2, q) = 0 \;.
\end{equation}

The quantity $\Phi_{\mu \nu}$ has the following tensor structure:
\begin{align}\label{tensor_mult}
\Phi_{\mu \nu}(q,k,m) &=  (- g_{\mu \nu}  q^2) \,
C(q^2,k^2,qk,m^2) +
k_{\mu} k_{\nu} \, D(q^2,k^2,qk,m^2) \nonumber \\
&+ (\mathrm{terms} \sim q_{\mu}, q_{\nu}) + (\mathrm{term} \sim
\varepsilon_{\mu \nu \alpha \beta} \, q^{\alpha} k^{\beta}) \;.
\end{align}
Due to condition~\eqref{partonometer_trans}, terms proportional to
$q_{\mu}$ or/and $q_{\mu}$ gives zero contribution to $N_{Ql}$
after convolution in Lorentz indices with the tensor $\Pi_{\mu
\nu}$.

\begin{figure}[htb]
\begin{center}
\epsfysize=3.5cm \epsffile{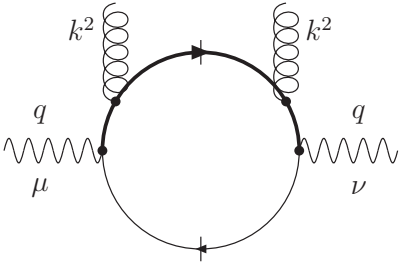}
\end{center}
\caption{The inclusive distribution of the massive gluon jet with
the virtuality $k^2$. The wavy line is the $W$ boson, whose
4-momentum is $q$. The thick quark line is a heavy quark, while
the thin line is a light quark. The cut quark lines mean that
these quarks are on-shell quarks.}
\label{fig:gluon_ladder_1}
\end{figure}

\begin{figure}[htb]
\begin{center}
\epsfysize=3.5cm \epsffile{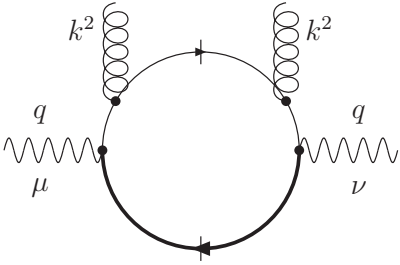}
\end{center}
\caption{The same as in Fig.~10, but with the gluon jet emitted by
the light quark.}
\label{fig:gluon_ladder_2}
\end{figure}

\begin{figure}[htb]
\begin{center}
\epsfysize=3.5cm \epsffile{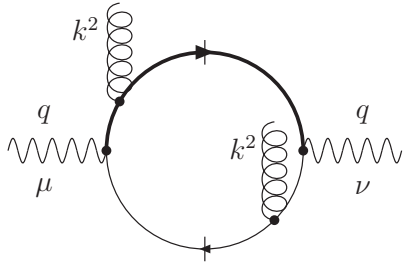}
\end{center}
\caption{The interference diagram which also contributes to the
inclusive distribution of the gluon jets with the virtuality $k^2$
in the $W$ boson.}
\label{fig:gluon_cross}
\end{figure}

In the first order in the strong coupling constant, $\Phi_{\mu
\nu}(q,k,m)$ is represented by the sum of three QCD diagrams
presented in Figs.~\ref{fig:gluon_ladder_1},
\ref{fig:gluon_ladder_2}, and \ref{fig:gluon_cross} (the crossed
diagram is taken with the factor 2).

One can use the following useful relation
\begin{align}\label{integration_tensor}
\int \!\! d^4k \, k_{\mu} k_{\nu} \, D(q^2,k^2,qk,m^2) &= (-
g_{\mu \nu} \, q^2) \, \int \!\! d^4k \, \frac{(qk)^2 - q^2
k^2}{3q^4}
\nonumber \\
&\times D(q^2,k^2,qk,m^2) + (\mathrm{term} \sim q_{\mu} q_{\nu} )
\;.
\end{align}
Detailed analytical calculations%
\footnote{Explicit analytic expressions for the functions
$C(q^2,k^2,qk,m^2)$ and $D(q^2,k^2,qk,m^2)$ are quite complicated
to be shown here. That is why, we will present only the final
expression \eqref{integrand_delE}.}
of the diagrams in
Figs.~\ref{fig:gluon_ladder_1}-\ref{fig:gluon_cross} show that
\begin{equation}\label{C_D_relation}
D = - \frac{q^4}{(qk)^2 - q^2 k^2} \, C \;,
\end{equation}
up to small power-like corrections of the type
$\mathrm{O}(m^2/m_W^2)$ and $\mathrm{O}(k^2/m_W^2)$. The
corrections $\mathrm{O}(k^2/m_W^2)$ can be neglected since it is
the region $k^2 \sim m^2$ that makes the leading contribution to
our main integral \eqref{delta_Ql} (see
Eq.~\eqref{large_negative_y} and a comment after it).

The normalization factor $N_0$ in \eqref{mult_0} is given by the
expression
\begin{align}\label{norm}
N_0 &= \frac{1}{3\pi} \, \Big( 1 - \frac{m^2}{m_W^2} \Big)^{2}
\Big( 1 + \frac{m^2}{2m_W^2} \Big) \nonumber \\
&\times \int \! \frac{d^4q}{(2\pi)^2} \, \delta (q^2 - m_W^2) \,
\delta ((p_1 + p_2 - q)^2 - m_W^2) \, (- g^{\mu \nu} q^2) \,
\Pi_{\mu \nu}
\nonumber \\
&\simeq \frac{1}{3\pi} \, \int \! \frac{d^4q}{(2\pi)^2} \, \delta
(q^2 - m_W^2) \, \delta ((p_1 + p_2 - q)^2 - m_W^2) \, (- g^{\mu
\nu} q^2) \, \Pi_{\mu \nu} \;.
\end{align}

Let us define
\begin{equation}\label{C_mod}
\tilde{C} = C \, \frac{\sqrt{(qk)^2 - m_W^2 k^2}}{(2\pi)^2 \, q^2}
\;.
\end{equation}
Then we derive from \eqref{mult_0}, and
\eqref{tensor_mult}-\eqref{norm}:%
\footnote{Note that the antisymmetric part of $\Phi_{\mu \nu}$
(i.e. the last term in Eq.~\eqref{tensor_mult}) gives zero
contribution to $\tilde{N}_{Ql}$.}
\begin{equation}\label{mult_1}
\tilde{N}_{Ql}(m_W^2, m^2) = \int\limits_{Q_0^2}^{(m_W - m)^2}
\frac{dk^2}{k^2} \, n_g(k^2) \int\limits_{(qk)_{-}}^{(qk)_{+}}
\!\! d(qk) \, \tilde{C}(m_W^2,k^2,qk,m^2) \;.
\end{equation}
The integration limits for the variable $qk$ are:
\begin{align}\label{limits_qk}
(qk)_{-} &= m_W \sqrt{k^2} \;,
\nonumber \\
(qk)_{+} &= (m_W^2 + k^2 - m^2)/2 \;.
\end{align}

The hadron multiplicity from the gluon jet with the \emph{fixed
virtuality} $k^2$ is related with the multiplicity from the gluon
jet whose \emph{virtuality varies up to} $k^2$~\cite{Kisselev:08}:
\begin{equation}\label{gluon_mult}
n_g(k^2) =  \Big( k^2 \, \frac{\partial}{\partial k^2} \Big)
N_g(k^2) \;.
\end{equation}

An explicit analytical expression for $\tilde{C}(m_W^2,k^2,qk)$
shows that
\begin{equation}\label{limits_C_mod}
\tilde{C}(m_W^2,k^2,qk,m^2)\Big|_{qk = (qk)_{-}} =
\tilde{C}(m_W^2,k^2,qk,m^2)\Big|_{qk = (qk)_{+}} = 0 \;.
\end{equation}

After introducing the notation
\begin{equation}\label{mult_final}
G(m_W^2,k^2,qk,m^2) = \Big( \!\! -k^2 \, \frac{\partial}{\partial
k^2} \Big) \, \tilde{C}(q^2,k^2,qk,m^2) \;,
\end{equation}
we come to the following equation:
\begin{equation}\label{mult_2}
\tilde{N}_{Ql}(m_W^2,m^2) = \int\limits_{Q_0^2}^{(m_W - m)^2}
\frac{dk^2}{k^2} \, N_g(k^2) \!\!
\int\limits_{(qk)_{-}}^{(qk)_{+}} \!\! d(qk) \,
G(m_W^2,k^2,qk,m^2) \;.
\end{equation}
Let us define
\begin{equation}\label{definition_E}
\int\limits_{(qk)_{-}}^{(qk)_{+}} \!\! d(qk) \,
G(m_W^2,k^2,qk,m^2) = C_F \, \frac{\alpha_s(k^2)}{\pi} \,
E_{Ql}(m_W^2, k^2,m^2) \;,
\end{equation}
where $C_F = (N_c^2 - 1)/(2N_c)$, and $N_c$ is a number of colors.
$E_{Ql}(m_W^2,k^2,m^2)$ is the inclusive spectrum of the gluon jet
emitted by $Q\bar{l}$-quark pair (see
Ref.~\cite{Petrov:95,Kisselev:07} for more details). Then the
multiplicity in heavy quark event \eqref{mult_2} can be
represented in the form:
\begin{equation}\label{mult_3}
N_{Ql}(m_W^2,m^2) = (n_Q +  n_l)+ C_F \!\!\!\!\!\!\!\!
\int\limits_{Q_0^2}^{(m_W - m)^2} \!\!\!\! \frac{dk^2}{k^2} \,
\frac{\alpha_s(k^2)}{\pi} \, N_g(k^2) \, E_{Ql}(m_W^2, k^2,m^2)
\;.
\end{equation}

Correspondingly, the hadron multiplicity  in the light quark event
is given by
\begin{equation}\label{mult_ll}
N_{l}(m_W^2) = 2n_l + C_F \! \int\limits_{Q_0^2}^{m_W^2} \!
\frac{dk^2}{k^2} \, \frac{\alpha_s(k^2)}{\pi} \, N_g(k^2) \,
E_{l}(m_W^2, k^2) \;,
\end{equation}
where $E_{l}(m_W^2, k^2) \equiv E_{Ql}(m_W^2, k^2,0)$. As a
result, we obtain the multiplicity difference:
\begin{equation}\label{delta_Ql}
\Delta_{Ql} = (n_Q - n_l) -  \!\int\limits_{Q_0^2}^{m_W^2} \!
\frac{dk^2}{k^2} \, N_g(k^2) \, \Delta E_{Ql}(k^2, m^2)  \;,
\end{equation}
where the following notation is introduced:
\begin{equation}\label{delE}
\Delta E_{Ql}(k^2, m^2) = E_{l}(m_W^2, k^2) - E_{Ql}(m_W^2,
 k^2,m^2) \;.
\end{equation}

Analytical calculations of the diagrams in
Figs.~\ref{fig:gluon_ladder_1}, \ref{fig:gluon_ladder_2}, and
\ref{fig:gluon_cross} result in the following expression:
\begin{align}\label{integrand_delE}
& G(q^2,k^2,qk,0) - G(q^2,k^2,qk,m^2)
\nonumber \\
&= C_F \, \frac{\alpha_s(k^2)}{\pi} \, \frac{8 m^2 \, qk \,
[k^2(q^2 - 2qk) + 2m^2(q^4 - 2q^2 qk + 2 (qk)^2) (qk)^2]}{q^2 \,
[k^2\, q^2\,(q^2 - 2qk) + 4m^2 (qk)^2]} \;.
\end{align}
As before, we have omitted small power-like corrections of the
type $\mathrm{O}(m^2/m_W^2)$ and $\mathrm{O}(k^2/m_W^2)$).%
\footnote{Remember that we have to put $q^2=m_W^2$ in
\eqref{integrand_delE}.}

One can conclude from Eqs.~\eqref{definition_E},
\eqref{integrand_delE} that $\Delta E_{Ql}(k^2, m^2)$ is a
function of the dimensional variable
\begin{equation}\label{variable_ro}
\rho = \frac{k^2}{m^2} \;.
\end{equation}
Indeed, let us define:
\begin{equation}\label{variable_change}
qk = [q^2(1 - x) + k^2 - m^2]/2 \;.
\end{equation}
The new variable $x$ varies within the limits:
\begin{equation}\label{limits_x}
0 \leqslant x \leqslant x_{\mathrm{max}} = \left( 1 -
\sqrt{k^2}/m_W \right)^2 - m^2/m_W^2  \;.
\end{equation}
One can safely set $x_{\mathrm{max}} = 1$, and obtain
\begin{align}\label{delta_E_Ql}
\Delta E_{Ql}(\rho) &=   \frac{1}{2} \, \int\limits_0^1 \!\! dx
(1-x)
\, \frac{(1-x)^2 \, (1 + x^2) + 4 \, x^2 \rho}{[(1-x)^2 + x \rho]^2} \nonumber \\
&= \frac{1}{4} \, [2 + \rho \, (3\rho - 2)] \ln \frac{1}{\rho} +
\frac{1}{4} \, (5 + 6\rho)
\nonumber \\
& +  \frac{1}{2} \, \rho \, (3\rho - 8) \, J(\rho) + 6 \, \frac{1
- J(\rho)}{\rho - 4} \;,
\end{align}\;.
where
\begin{equation}\label{J}
J(\rho) =
  \begin{cases}
     \sqrt{\frac{\rho}{\rho - 4}}
     \ln \left( \frac{\sqrt{\rho} \, + \sqrt{\rho - 4}}{2}
     \right),
     & \rho > 4 \;, \cr
     \ 1 \;, & \rho = 4 \;, \cr
     \sqrt{\frac{\rho}{4 - \rho}}
     \arctan \left( \frac{\sqrt{4 - \rho}}{\rho} \right),
     & \rho < 4 \; .
  \end{cases}
\end{equation}

Remember that we considered the process $e^-e^+ \rightarrow W^+
W^-$ and compared two possible subsequent hadronic decays of one
of the bosons, $W^+ \rightarrow c \bar{d} \ (c\bar{s}) +
\mathrm{(gluon \ jets)}$, and $W^+ \rightarrow u \bar{d} \
(u\bar{s}) + \mathrm{(gluon \ jets)}$. The quantity $\Delta
E_{cl}(k^2/m_c^2)$ \eqref{delta_E_Ql} describes the difference of
the distributions of the gluon jets in their invariant mass $k^2$
in these processes ($m_c$ is the mass of the charm quark). Let us
stress that Eq.~\eqref{delta_E_Ql} coincides with the expression
for the difference of the gluon jet distributions derived in the
case when the $W^+$ boson is a product of the top weak decay $t
\rightarrow b + W^+$ (see Eq.~(35) from Ref.~\cite{Kisselev:top}).

\begin{figure}[htb]
\begin{center}
\epsfysize=6cm \epsffile{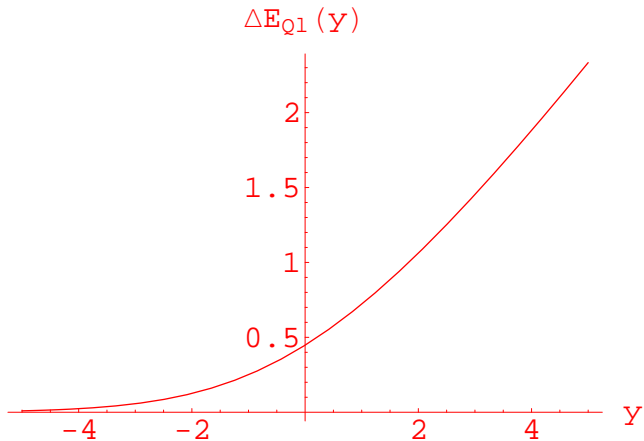}
\end{center}
\caption{The function $\Delta E_{Ql}(y)$.}
\label{fig:delta_E_Ql}
\end{figure}

In terms of variables
\begin{equation}\label{y}
y = \ln \frac{m^2}{k^2} \equiv \ln \frac{1}{\rho} \;,
\end{equation}
and
\begin{equation}\label{Y_Q}
Y = \ln \frac{m^2}{Q_0^2} \;,
\end{equation}
the multiplicity difference looks like
\begin{equation}\label{delta_N_Ql}
\Delta_{Ql} = (n_Q - n_l) -  \!\int\limits_{-\infty}^{Y} \! dy \,
N_g(Y - y) \, \Delta E_{Ql}(y)  \;,
\end{equation}
with $E_{Ql}$ defined by Eqs.~\eqref{delta_E_Ql}, \eqref{J}. The
function $\Delta E_{Ql}(y)$ is shown in Fig.~\ref{fig:delta_E_Ql}.

Since
\begin{equation}\label{large_negative_y}
\Delta E_Q(y)\Big|_{y \rightarrow -\infty} \simeq \frac{11}{3} \,
e^{-|y|} \;,
\end{equation}
the integral in \eqref{delta_N_Ql} converges rapidly at the lower
limit. Asymptotics of $\Delta E_{Ql}(y)$ at large $y$ is the
following:
\begin{equation}\label{large_positive_y}
\Delta E_{Ql} (y)\Big|_{y \rightarrow \infty} \simeq \frac{1}{2}
\, \Big( y - \frac{1}{2}  \Big) \;.
\end{equation}

In Ref.~\cite{Petrov:95} the analogous formula for the
multiplicity difference in $e^+e^-$ events \emph{not mediated} by
the $W$ bosons  \eqref{delta_Ol_def} was derived:
\begin{equation}\label{delta_N_Q}
\delta_{Ql} = 2(n_Q - n_l) -  \!\int\limits_{-\infty}^{Y} \! dy \,
N_g(Y - y) \, \Delta E_{Q}(y)  \;,
\end{equation}
where
\begin{align}\label{Delta_E_Q}
\Delta E_{Q}(\rho) & =  1 + \rho \, \Big( \frac{7}{2}\rho - 3
\Big) \ln \frac{1}{\rho} + \Big( \frac{9}{2} + 7\rho \Big)
\nonumber \\
& +  \rho \, (7\rho - 20) \, J(\rho) + 20 \, \frac{1 -
J(\rho)}{\rho - 4} \;,
\end{align}
with $J(\rho)$ defined above \eqref{J}.

\begin{figure}[htb]
\begin{center}
\epsfysize=6cm \epsffile{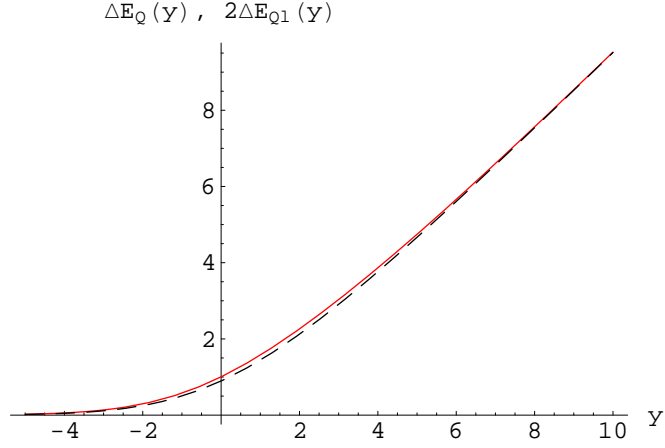}
\end{center}
\caption{The function $\Delta E_Q(y)$ (solid curve) vs. function
$2\Delta E_{Ql}(y)$ (dashed curve).}
\label{fig:dE_Q_dE_Ql}
\end{figure}

The function $\Delta E_Q(y)$ has the following asymptotics at $y
\rightarrow \infty$:
\begin{equation}\label{small_rho}
\Delta E_Q(y)\Big|_{y \rightarrow \infty} \simeq  y - \frac{1}{2}
\;.
\end{equation}
As one can see from Eqs.~\eqref{small_rho},
\eqref{large_positive_y}, $\Delta E_{Ql}(y) =\Delta E_Q(y)/2$ at
large $y$. Moreover, numerical calculations show that $\Delta
E_{Ql}$ is very close to $\Delta E_Q/2$ \emph{at all} $y$ (see
Fig.~\ref{fig:dE_Q_dE_Ql}).
Thus, we get the prediction:
\begin{equation}\label{Delta_vs_delta}
\Delta_{Ql} =  \frac{1}{2} \, \delta_{Ql} \;.
\end{equation}

Now we can calculate the total hadron multiplicity from the $W$
boson:
\begin{align}\label{n_W}
N_W &= N_{l}(m_W) +  \frac{|V_{cd}|^2 + |V_{cs}|^2}{|V_{ud}|^2 +
|V_{us}|^2 + |V_{cd}|^2 + |V_{cs}|^2} \, (N_{cl} - N_l)
\nonumber \\
&= N_{l}(m_W) + \frac{1}{4} \, \delta_{cl} \;.
\end{align}
For our numerical estimates, we will use the corrected
experimental value of $\delta_{cl}$ from
Ref.~\cite{Dokshitzer:06}:
\begin{equation}\label{delta_cl_exp}
\delta_{cl}= 1.03 \pm 0.34
\end{equation}
The light quark multiplicity at the energy $W=m_W$ was recently
estimated to be~\cite{Kisselev:top}:
\begin{equation}\label{light_quark_mult}
N_{l}(m_W) = 19.09 \pm 0.18 \;,
\end{equation}
that results in
\begin{equation}\label{n_W_num}
N_W = 19.34 \pm 0.21 \;.
\end{equation}

Let us estimate effects associated with W boson decays into $u
\bar{b}$ and $c \bar{b}$-pairs. The account of the $b$-quark
production increase the multiplicity from the $W$ boson
\eqref{n_W} by the quantity $\Delta N_W^b$, i.e. $N_W \rightarrow
N_W + \Delta N_W^b$, where
\begin{equation}\label{b_mult}
\Delta N_W^b = |V_{ub}|^2 \, (N_{bl} - N_l) +  |V_{cb}|^2 \,
(N_{bc} - N_l) \;.
\end{equation}
The first term in \eqref{b_mult} is equal to
\begin{equation}\label{bu_mult}
\frac{1}{2} \, |V_{ub}|^2 \, \delta_{bl} \;.
\end{equation}
In order to estimate the second term, we will use the fact that
the emission of the massive gluon jets by heavier quark is
suppressed, that results in $E_{bc} < E_{bl}$. Thus, we get the
inequality:
\begin{equation}\label{bc_mult}
N_{bc} - N_l < N_{bl} - N_l = (n_c - n_l) + \frac{1}{2} \,
\delta_{bl} \;.
\end{equation}
Using the average values of $n_c = 2.6$, $n_l = 1.2$, and
$\delta_{bl} = 3.12$~\cite{Dokshitzer:06}, as well as CKM matrix
elements $|V_{ub}| = 3.95 \cdot 10^{-3}$, $|V_{cb}| = 38.6 \cdot
10^{-3}$~\cite{PDG} we obtain from \eqref{b_mult}-\eqref{bc_mult}
that
\begin{equation}\label{b_mult_num}
\Delta N_W^b < 1.2 \cdot 10^{-2} \;.
\end{equation}
It means that one can ignore the contribution to $N_W$ from the
multiplicity difference between $ub$($cb$)-event and light quark
event.

Now let us take into account that the $W$ boson with the 4-momenta
$q$ in Fig.~\ref{fig:W_emission} is not on-shell, but has a
distribution in its invariant mass $q^2$. The denominator of the
$W$ boson propagator is equal to
\begin{equation}\label{W_width}
q^2 - m_W^2 + i m_W \Gamma_W \;,
\end{equation}
where $\Gamma_W$ is the full $W$ width. As a result, the hadron
multiplicity from the $W$ boson is given by the formula:
\begin{equation}\label{n_W_corr}
N_W  - \frac{1}{4} \, \delta_{cl} =  \frac{1}{H}  \!\!\!
\int\limits_{Q_0^2}^{(W - m_W)^2} \!\!\!\!\! dq^2 \, \frac{1}{(q^2
- m_W^2)^2 + m_W^2 \Gamma_W^2} \, F(W^2, q^2) \, N_l(q^2) \;,
\end{equation}
where
\begin{equation}\label{norm_corr}
H = \!\!\!\! \int\limits_{Q_0^2}^{(W - m_W)^2} \!\!\!\!\! dq^2 \,
\frac{1}{(q^2 - m_W^2)^2 + m_W^2 \Gamma_W^2} \, F(W^2,q^2) \;.
\end{equation}
The function $F(W^2,q^2)$ is defined by the invariant part of the
diagram in Fig.~\ref{fig:W_emission} integrated in $d^3q$ at fixed
$q^2$. It can be chosen to be dimensionless,%
\footnote{Since $q^2$-independent dimensional factors may be
safely omitted in $F$.}
i.e. dependent on the ratio $q^2/W^2$.

The main contributions to the integrals in \eqref{n_W_corr},
\eqref{norm_corr} come from the region $q^2 - m_W^2 \sim m_W
\Gamma_W$. The argument of the function $F(q^2/W^2)$ varies rather
slowly in this region, contrary to the factor $[(q^2 - m_W^2)^2 +
m_W^2 \Gamma_W^2]^{-1}$ which has a sharp peak around the point
$q^2 = m_W^2$. Thus, we can assume that an explicit form of
$F(q^2/W^2)$ is not important and put $F = 1$ in
Eqs.~\eqref{n_W_corr}, \eqref{norm_corr}. By using a fit of the
data on light quark multiplicity $N_l(E^2)$, we can estimate a
possible variation of the multiplicity \eqref{n_W_corr}. The
numerical calculations show that at $W = 183$ GeV (collision
energy of LEPII experiments on $W^+W^-$
production~\cite{OPAL}-\cite{DELPHI}) it differs from $N_l(m_W)$
by the quantity $0.05$.%
\footnote{This difference becomes smaller with the increase of
$W$.}
Then we sum the uncertainty $2 \times 0.05 = 0.10$ with the errors
of $N_W$ \eqref{n_W_num} in quadrature and obtain:
\begin{equation}\label{n_W_corr_num}
N_{l}(m_W) = 19.34 \pm 0.23 \;.
\end{equation}

Our result \eqref{n_W_corr_num} is in a good agreement with the
experimental data from OPAL~\cite{OPAL},
\begin{align}
N_W &= 19.3 \pm 0.3 \pm 0.3 \;, \label{OPAL} \intertext{and
DELPHI~\cite{DELPHI},} N_W &= 19.44 \pm 0.13 \pm 0.12 \;.
\label{DELPHI}
\end{align}

Formulae \eqref{Delta_vs_delta} and \eqref{n_W} is our main
theoretical result. Let us underline that these equations
\emph{relate measurable quantities}, and they \emph{do not depend
on the explicit form} of the hadron multiplicity from the gluon
jet $N_g(k^2)$.

Our main prediction is
\begin{equation}\label{Delta_num}
\Delta_{cl} = N_{cl} - N_l = 0.52 \pm 0.17 \;.
\end{equation}
This prediction can be checked by using available LEPII data on
hadron multiplicities in $W^+W^- \rightarrow q\bar{q}' \, l
\bar{\nu}_l$ and  $W^+W^- \rightarrow q\bar{q}' \, q\bar{q}'$
events~\cite{OPAL,DELPHI} as well as future data from the ILC. It
will be one more experimental test of the universality (i.e. the
process-independent character) of the QCD evolution in the
multiple hadron production.

Both the multiplicity \eqref{n_W_corr_num}, and multiplicity
difference \eqref{Delta_num} can be also measured at the LHC. One
of the processes to look for mass-dependent effects in the QCD
evolution is a \emph{single} $W$ production with its subsequent
decay into hadrons. Another possibility is a production of
\emph{two} $W$'s, one of which decays in a lepton mode, while the
other goes into hadrons with (without) charm production.



\section{Conclusions}

In the present paper we have calculated the hadron multiplicity
from $W$ boson in pQCD. The difference of the hadron
multiplicities in the $W$-boson hadronic decays \emph{with} and
\emph{without} charm production is also estimated.
Previously~\cite{Petrov:95}-\cite{Kisselev:08} the multiplicity
difference in $b\bar{b}$ ($c\bar{c}$)-event and light quark event
was calculated on the same ground. The nice agreement of these
theoretical predictions with the LEP data confirms the universal
character of the mechanism of multiple hadron production in QCD
via evolution of quark-gluon showers.




\end{document}